\documentstyle[12pt,epsf]{article}
\newcounter{pepe}                         
{\end{eqnarray}%
\setcounter{equation}{\arabic{pepe}}%
} 

\newcommand{\Rb}{{\cal   R}}

\newcommand{\nne}{{\bf  e}}   
\newcommand{\nj}{{\bf   j}}
\newcommand{\nk}{{\bf   k}}
\newcommand{\np}{{\bf   p}}       
\newcommand{\nq}{{\bf   q}}
\newcommand{\nr}{{\bf   r}}

\newcommand{\nv}{{\bf   v}}

\newcommand{\hp}{{\bf   \hat{p}}}
\newcommand{\hr}{{\bf   \hat{r}}} 

\newcommand{\nOmega}{\mbox{\boldmath$\Omega$}} 
\newcommand{\hphi}{\hat{\mbox{\boldmath$\phi$}}}

\newcommand{\cdr}{\framebox[1cm]{\rule[-5mm]{0cm}{1cm}}}
\newcommand{\gr}{$^{\rm o}$}

\begin{document}
\begin{titlepage}
\mbox{} 
\vspace*{2.5\fill} 

{\Large\bf 
\begin{center}
%
Final-State Interactions in $(e,e'p)$ Reactions \\
with Polarized Nuclei
%
\end{center}
} 

\vspace{1\fill} 

\begin{center}
{\large J.E. Amaro$^{1}$ and T.W. Donnelly$^2$} 
\end{center}

{\small 
\begin{center}
$^1$ {\em Departamento de F\'{\i}sica Moderna, 
          Universidad de Granada,
          Granada 18071, Spain}  \\   
$^2$ {\em Center for  Theoretical  Physics,
          Laboratory   for Nuclear Science  and  
          Dept. of Physics, }\\   
     {\em Massachusetts Institute  of  Technology, 
          Cambridge,  MA  02139, U.S.A.}\\[2mm] 
\end{center}
} 

\kern 1.5 cm 

\hrule \kern 3mm 

{\small 
\noindent
{\bf Abstract} 
\vspace{3mm} 

The cross section for coincidence, quasielastic proton knock-out
by electrons from a polarized $^{39}$K nucleus is computed in DWIA
using an optical potential in describing the wave function
of the ejected nucleon. The dependence of the FSI on the initial
polarization angles of the nucleus is analyzed and explained
in a new, semi-classical picture of the reaction in which the nuclear 
transparency decreases as a function of the amount of nuclear matter
that the proton has to cross, thus providing a method for obtaining 
detailed information on its 
mean free path in finite nuclei. We propose a procedure to find
the best initial kinematical conditions for
minimizing the FSI which will be useful as a guide for future experiments  
with polarized nuclei.
 } 

\kern 3mm \hrule 

\kern 4cm



\noindent MIT/CTP\#2784 \hfill October 1998

\end{titlepage}


\section{Introduction}


The interpretation of many experiments in nuclear physics
requires an understanding of how nucleons propagate through nuclei
\cite{Gee89,Gar92}. In particular, in quasi-free coincidence 
$(e,e'p)$ reactions, an electron transfers an energy $\omega$ to a 
nucleon in the nucleus. This high-energy nucleon exits from the nucleus,
leaving the daughter nucleus in a state with definite energy and momentum.
The main effect of the final-state interaction (FSI) felt by the ejected
nucleon as it travels across the residual nucleus is a reduction of the 
cross section due to attenuation caused by interactions with the nuclear matter 
through which the nucleon must propagate \cite{Fru84}. 
This attenuation can be described in terms of a complex one-body 
optical potential, whose imaginary part accounts for 
the loss of flux produced by transitions to channels other than the 
elastic one. Accordingly, it is expected that the so-called 
distorted wave  impulse approximation (DWIA) yields a cross section 
which is smaller than in the plane-wave impulse approximation (PWIA), and
hence that the nuclear transparency, defined as the ratio between the DWIA
and PWIA cross sections, is usually less than one. This reduction is
observed in experiments for low \cite{Fru84}, moderate 
\cite{Gee89,Gar92}  and high momentum transfer \cite{Mak94,ONe95}, and
at present there is no unambiguous evidence of a restoration of the
full nuclear transparency (i.e., a reduction of the FSI) by effects
such as those that underlie the idea of color transparency \cite{Bla93}.

 In this paper we explore the effects of the FSI in $(e,e'p)$ 
reactions using polarized nuclei. 
The importance of having the FSI under control is clear
if one is trying to study nuclear properties (such as initial-state 
momentum distributions) or reaction mechanisms (such as specifics of 
the nuclear electromagnetic current).
All of the measurements mentioned above involving medium and heavy nuclei have 
been performed with unpolarized targets and hence only the global
effects of FSI averaged over all polarization directions have been addressed
experimentally to date. Using instead polarized nuclei as targets, 
new possibilities to extract the full tri-dimensional momentum distribution 
of nuclei will become available \cite{Cab93,Cab94,Gar95,Cab95}; of course, 
as in the unpolarized case, the FSI also enter here and must be taken 
into account.

In fact, as established in a previous paper \cite{Ama98}, the various
multipole reduced responses that enter as components of the cross section
show different sensitivities to the FSI. Hence it is expected that there should
be a dependence of FSI effects --- or nuclear transparency --- on the choice
of polarization angles. There are only a few preliminary studies 
of $(e,e'p)$ reactions involving polarized, medium and heavy nuclei in 
DWIA \cite{Gar95,Bof88}, and these report only a few examples for 
illustrative purposes without 
developing detailed insights into the roles played by FSI
for different polarizations. 

Hence a theoretical study of such 
effects is motivated when attempting to understand the new issues of 
nuclear transparency as a function of the orientation of the nuclear spin.
The goal of the present work is to show how the variations of the 
transparency can be understood in terms of the orientation of 
the initial-state nucleon's orbit (the one that in our model primarily
carries the nuclear polarization) and of the attenuation of the ejected 
nucleon's flux through its dependence on the length of the path that it
travels in the nucleus. Although the method we follow is developed 
here for the particular case of the ejection of a proton from the 
$d_{3/2}$ shell of polarized $^{39}$K leaving $^{38}$Ar$_{g.s.}$ as
the daughter nucleus, it can be generalized for use with any polarized
nucleus and can be addressed using more sophisticated nuclear models. 
The present choice is, however, prototypical. Using it we shall show 
that one is able to predict 
the orientations of the target polarization that are optimal for 
minimizing the FSI effects, that is, to make the nucleus as ``transparent'' 
as possible. As in these cases the  FSI effects are minimized, 
they provide the ideal situations to study other issues such as specifics 
of initial-state nuclear structure. As we shall show below,
this special situation occurs when the nucleon is ejected directly away 
from the nuclear surface. On the other hand,  when the nucleon is ejected 
from the nuclear surface but in the opposite direction --- into the nucleus 
--- it has to cross the entire nucleus to exit on the opposite side, 
and the FSI effects are then found to be maximal (that is, one has the minimum 
transparency). This second situation is ideal for detailed studies of 
the imaginary, absorptive part of the FSI. Finally,  intermediate 
situations arise in which the nucleon is ejected from the surface 
in a direction roughly ``tangent'' to the nuclear surface, and there 
the re-scattering mechanisms that originate with the real and spin-orbit 
parts of the potential have a major influence. All of these situations 
can be selected simply by changing the direction of the nuclear polarization.
If one focuses on the first and second situations, but not the third
in which re-scattering is appreciable, it is possible to parametrize the 
transparency in terms of a  mean free path for {\em finite} nuclei.

The organization of the present work is as follows: in sect.~2 we review the
details of the formalism for describing $(e,e'p)$ reactions with 
polarized nuclei which are of relevance for the discussions to follow, 
and present our model with some of the details of the calculation. 
In sect.~3 we show the numerical results of the present calculation of 
the cross section for different nuclear polarizations. In sect.~4 we 
introduce a semi-classical model of the reaction in order to provide 
a physical picture of the dependence presented in sect.~3 of the nuclear 
transparency as a function of the polarization angles
and to parametrize it in terms of an effective mean free path. 
Finally, in sect.~5 we draw our conclusions.


\section{Coincidence cross section 
         of polarized nuclei}


\subsection{Kinematics and cross section}

First we introduce the definitions of the kinematics we use. An electron 
is scattered by a nucleus $|A\rangle$ of mass $M_A$, transferring to it 
an energy $\omega=E_e-E'_e$ and momentum $\nq=\nk_e-{\nk}'_e$.
In the final state a proton of mass $M$ with momentum $\np'$ and energy
$E'$ is detected in coincidence with the electron. The daughter nucleus 
is left in a definite state $|B\rangle$ with mass $M_B^*$, which we consider 
to be located in the discrete spectrum; accordingly we integrate over 
the missing energy 
 \begin{eqnarray}
   E_m &=& \sqrt{(\omega+M_A-E')^2-(\nq-\np')^2}-M_B \nonumber\\
&=& \sqrt{E_B^2-p_B^2} -M_B \equiv M_B^*-M_B,
\end{eqnarray}
in order to select one of the discrete final states for $B$ in the 
missing-energy spectrum. Here $M_B$ is the daughter ground-state mass.
Thus $E_B=\omega+M_A-E'$ is the total energy of the residual nucleus, 
$\np= -\np_B= \np'-\nq$ 
is the missing momentum, and $M_B^*=\sqrt{E_B^2-p_B^2}$ is the mass of
the daughter nucleus in its (in general) excited state; we neglect recoil in the
present work.

In addition, we consider the initial nucleus to be 100\% polarized 
in a direction given by the unit vector $\nOmega^*$; that is, the 
initial hadronic state is labeled
\begin{equation}
|A \rangle = |A(\Omega^*)\rangle = R(\Omega^*)|J_iJ_i\rangle ,
\end{equation}
where $R(\Omega^*)$ is a rotation operator which maps the $z$-axis
(the {\bf q}-direction) 
onto the  $\nOmega^*$ direction and $J_i$ is the total spin
of the nucleus $A$. For simplicity in the following arguments,
in this work we do not consider polarized electrons, since only the
polarization of the nucleus is essential for the model presented in sect.~4;
also in the present work the polarization of the final state is assumed 
not to be specified.

Assuming plane waves for the electrons and working in a reference
system where the $z$-axis points in the positive $\nq$ direction 
and the $x$-axis is in the electron scattering plane, one has the following 
expression for the cross section \cite{Ras89}:
\begin{equation}
\Sigma \equiv \frac{d\sigma}{dE'_ed\Omega'_ed\Omega'}
       = \sigma_M\left( v_L\Rb^L + v_T\Rb^T + 
                        v_{TL}\Rb^{TL} + v_{TT}\Rb^{TT} 
                 \right),
\end{equation} 
where $\sigma_M$ is the Mott cross section, $v_K$ are the electron 
kinematical factors given in \cite{Ras89}, and $\Rb^K$ are the
nuclear response functions, which are given as (real) linear
combinations of components of the hadronic tensor
\begin{equation}
W^{\mu\nu} = \sum_{m_s m_B}
             \langle \np' m_s B | J^{\mu}(\nq,\omega)|A\rangle^*
             \langle \np' m_s B | J^{\nu}(\nq,\omega)|A\rangle,
\end{equation}
with $J^{\mu}(\nq,\omega)$ the nuclear electromagnetic current
operator,
and $m_s$, $m_B$ are the (undetected) magnetic quantum numbers of the
final unpolarized hadrons.

\subsection{Nuclear and reaction models}

We now give some details concerning the nuclear model of the reaction 
used to compute the cross section from polarized
$^{39}$K. For a more complete description of the model see references
\cite{Ama98,Ama96a,Ama96b}.
We assume that the ground state is described as a hole ($h$) in the
$d_{3/2}$ shell of $^{40}$Ca:
\begin{equation}
|A\rangle = b^{\dagger}_h\;|^{40}{\rm Ca}\rangle.
\end{equation}
The ground state of the daughter nucleus $^{38}$Ar is
described as two holes in the $d_{3/2}$ shell of 
$^{40}$Ca, coupled to final spin $J_B=0$:
\begin{equation}
|B\rangle = [b^{\dagger}_hb^{\dagger}_h]_0\;|^{40}{\rm Ca}\rangle.
\end{equation}
Thus we consider the case where the proton is ejected from the outer
shell of $^{39}$K.
The wave function for the hole state is obtained by solving the
Schr\"odinger equation with a Woods-Saxon potential \cite{Ama94}.

As explained in Ref. \cite{Ama98}, the case of a hole nucleus is not as simple
as the reverse situation of a nucleus with just a single particle in the outer
valence shell, for in this latter case the daughter nucleus always has $J_B=0$
for a particle ejected from that outer shell.
In the case of a hole, the residual nucleus can be in several states
with different spins, and, as a result, the response $\Rb^K$
of a one-hole nucleus is {\em not} the same as that for a 
one-particle nucleus. Yet in Ref. \cite{Ama98}
it is shown that in the particular case  where $J_B=0$ they are proportional,
with a factor $2/(2j_h+1)$ (see eq. (82) in the above-cited reference).  
In the case of interest here, with $j_h=3/2$, that factor is equal to $1/2$.

Concerning the ejected particle wave function, in the present 
treatment it is obtained by solving the Schr\"odinger equation 
using a complex optical potential fitted to elastic proton scattering
from a variety of nuclei \cite{Sch82}. The partial wave $(l,j)$ of
the outgoing proton with wave number $p'$ is normalized through the 
asymptotic behaviour
\begin{equation}
R_{lj}(r) 
\sim \sqrt{\frac{2M}{\pi\hbar^2p'}}
     {\rm e}^{-i(\sigma_l+\delta_{lj}^*)}
      \sin\left( p'r -\eta\log 2p'r-l\frac{\pi}{2}+\sigma_l+\delta_{lj}^*
          \right).
\end{equation}
This condition reflects the boundary condition (neglecting the Coulomb 
potential)
\begin{equation}
R_{lj}(r) \sim S_{lj}^*{\rm e}^{-ip'r}-{\rm e}^{ip'r} ,
\end{equation}
where 
\begin{equation}
S_{lj}={\rm e}^{2i\delta_{lj}}=
\eta_{lj}{\rm e}^{2i{\rm Re}\delta_{lj}}
\end{equation}
is the $S$-matrix partial-wave amplitude and $\eta_{lj}\le 1$, corresponding
to an absorptive potential.


Given the value $t'\equiv E'-M = \epsilon_{\ell j}+\omega$ of the kinetic 
energy of the ejected nucleon, where $\epsilon_{\ell j}$ is the
initial-state (bound, $\epsilon_{\ell j}<0$) nucleon's eigenvalue, the momentum $p'$ 
is computed using relativistic kinematics 
\begin{equation}
p'=\sqrt{E'^2-M^2}=\sqrt{2Mt'\left(1+\frac{t'}{2M}\right)},
\end{equation}
which is equivalent to making the substitution 
$t'\rightarrow t'(1+\frac{t'}{2M})$
in a non-relativistic model. In addition, we use a relativized 
electromagnetic current which is appropriate for electron scattering
calculations near the quasielastic peak \cite{Ama96a,Ama98b}.
In this way, although based in a non-relativistic approach, our
model retains many aspects of relativity which allow us to apply it
for high momentum transfers.


We compute the current matrix elements by performing a multipole
expansion of the current operator. Accordingly, we write the 
final and initial nuclear wave functions as sums of multipoles
of the ejection angles and polarization angles, respectively.
The sums over multipoles of the current and final states are infinite, and 
thus we need to truncate the expansion for values where convergence is reached.
As a test of the convergence we perform the calculation of the PWIA
responses in two different ways --- with the multipole expansion 
and with the factorized expressions \cite{Cab93} --- allowing us to 
fix the number of multipoles needed. From our study of response functions
\cite{Ama98} for momenta $q\le 700$ MeV/c, we have found that it is
enough to sum up to $J=32$ in the multipole expansion for $^{39}$K.

\section{Dependence of the cross section and FSI effects on the
            nuclear polarization}

Next we present the numerical results of our calculations. 
In order to show the variety of FSI effects in the cross section,
we have performed a calculation for different values of the 
angles $\theta^*$ and $\Delta\phi=\phi-\phi^*$. 
Here $\Omega^*=(\theta^*,\phi^*)$ are the polarization angles, 
or spherical coordinates of the vector $\nOmega^*$ in a coordinate
system with $\nq$ in the $z$-axis and with the $xz$-plane as the
electron scattering plane; $\phi$ is the azimuthal angle of the plane in which
the ejected proton lies in this reference system. 

\begin{table}[ht]
\begin{center}
\begin{tabular}{ccccc@{\hspace{1cm}}cccc}
$\theta^*$   &      &      &      &      &      &      &      &      \\
$\downarrow$ &      &      &      &      &      &      &      &      \\
$0^{\rm o}$  & \multicolumn{4}{c}{\cdr}  &      &      &      &      \\[7mm]
$45^{\rm o}$  & \cdr & \cdr & \cdr & \cdr & \cdr & \cdr & \cdr & \cdr \\[7mm]
$90^{\rm o}$  & \cdr & \cdr & \cdr & \cdr & \cdr & \cdr & \cdr & \cdr \\[7mm]
$135^{\rm o}$ & \cdr & \cdr & \cdr & \cdr & \cdr & \cdr & \cdr & \cdr \\[7mm]
$180^{\rm o}$ & \multicolumn{4}{c}{\cdr}  &      &      &      &      \\[7mm]
$\Delta\phi\rightarrow$
               &0\gr &45\gr &90\gr &135\gr&180\gr&225\gr&270\gr&315\gr 
\end{tabular}
\end{center}
{\small \sf
{\bf Table  1:} Polarization angles for the various panels given in
         figs.~1 and 2
}
\end{table}

The results are shown in figs.~1 and 2 for kinematics
corresponding to the quasielastic peak and in-plane emission
\begin{equation}
q=500\; {\rm MeV}/c, \kern 5mm
\omega= 133.5\; {\rm MeV}, \kern 5mm
\phi=0, \kern 5mm
\theta_e = 30^{\rm o} ,
\end{equation}
where $\theta_e$ is the electron scattering angle. Note that for 
these particular kinematics, $\phi^*=-\Delta\phi$.
In table 1 we give the key for obtaining the polarization 
angles used in  each panel in figs.~1 and 2.

In fig.~1 we show 14 panels corresponding to different
polarizations $\Omega^*$. The ones at the top and bottom 
correspond to $\theta^*=0$ and 180$^{\rm o}$ respectively. 
The remaining 12 panels
correspond, from top to bottom, to $\theta^*=45,90,135^{\rm o}$
and, from left to right, to $\Delta\phi=0,45,90,135^{\rm o}$.
The  solid lines in  this figure represent values of the 
cross section $\Sigma$ as a function of the missing momentum $p$
in DWIA for different polarizations contained in half of the sphere
in $\nOmega^*$-space; the top and bottom panels correspond to the 
north and south poles of the sphere.
The other half of the sphere is represented in fig.~2, where again, 
from top to bottom, $\theta^*=45,90,135^{\rm o}$,
and, from left to right,  $\Delta\phi=180,225,270,315^{\rm o}$.
In these figures the dashed lines are the cross sections computed 
in PWIA, i.e., without FSI. The dotted lines correspond to the DWIA,
but including in the FSI just 
the central imaginary part of the optical potential,
while the dash-dotted lines include in addition the central real part
of the potential. 

\begin{figure}[hptb]
\begin{center}
\leavevmode
\epsfbox[100 250 502 700]{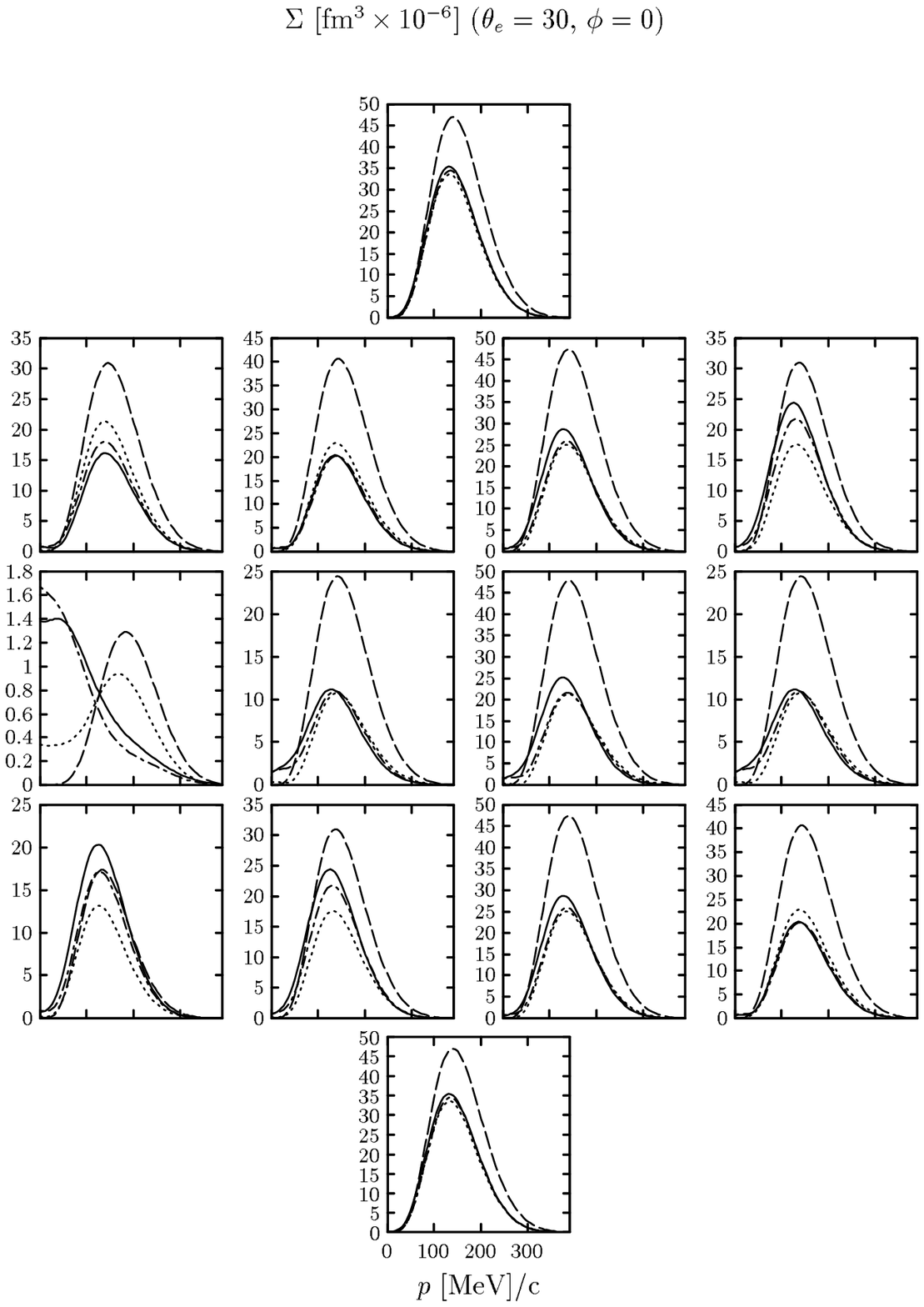}
\end{center}
{\small \sf
{\bf Figure 1:} Cross section computed for different values of 
the nuclear polarization angles as shown in the first half 
of table 1. From top to bottom:
$\theta^*=0,45,90,135,180^{\rm o}$. From left to right 
$\Delta\phi=0,45,90,135^{\rm o}$.  The meaning of the curves  
is the following: solid: DWIA; dashed: PWIA; dotted: DWIA 
but with just the imaginary part of the central optical potential;
dash-dotted: DWIA without spin-orbit contributions.
}
\end{figure}

\begin{figure}[htbp]
\begin{center}
\leavevmode
\epsfbox[100 390 502 700]{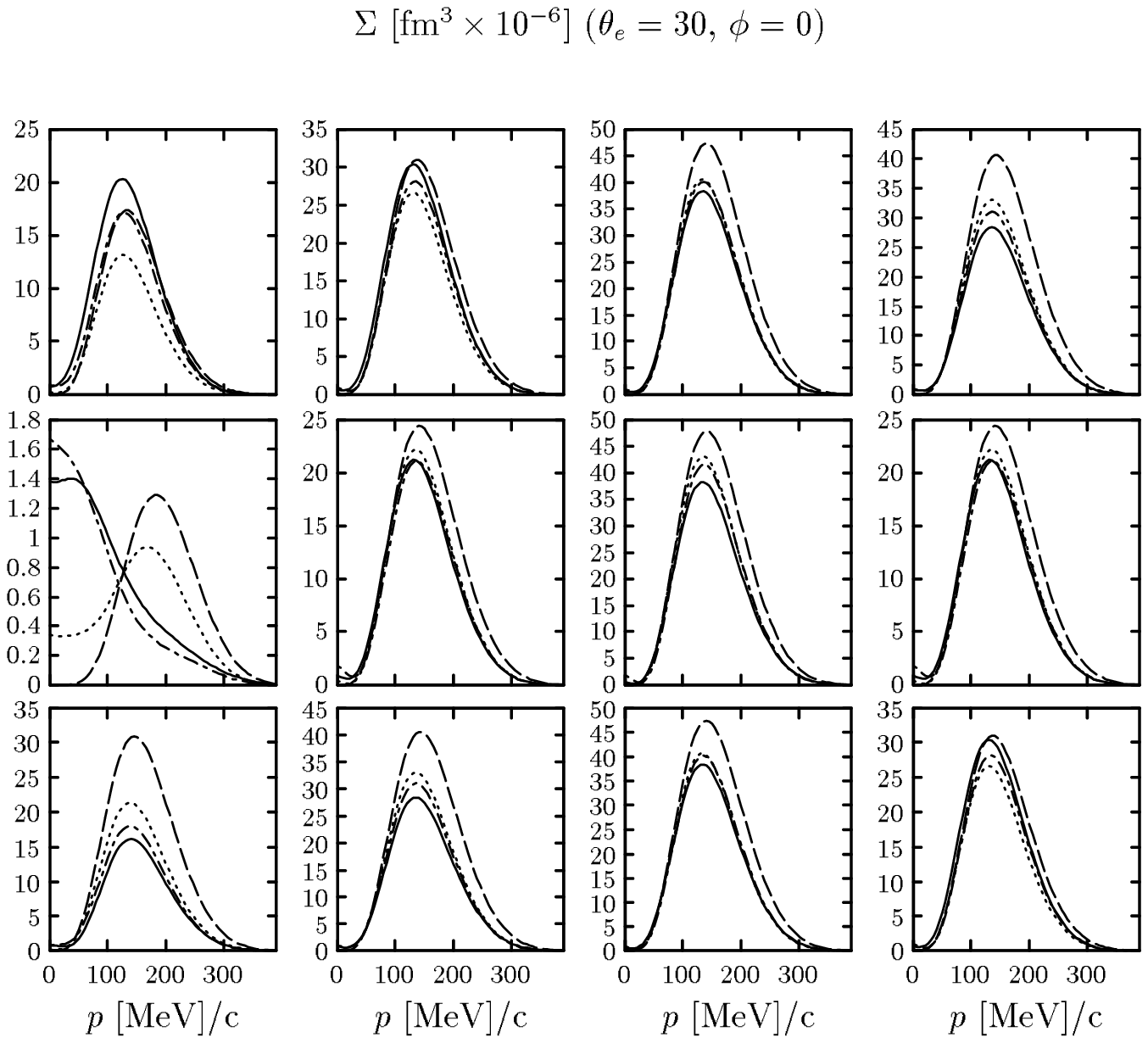}
\end{center}
{\small \sf
{\bf Figure 2:} Cross section computed for different values of 
the nuclear polarization angles as shown in the second half of 
table 1. From top to bottom:
$\theta^*=45,90,135^{\rm o}$. From left to right 
$\Delta\phi=180,225,270,315^{\rm o}$.
The meaning of the curves  is the same as in fig.~1.
}
\end{figure}
Looking at the behaviour of $\Sigma$ versus the angles
($\theta^*,\Delta\phi$) across 
all of the 26 panels of figs.~1--2, we see that the cross section and the 
relative FSI effect --- i.e., the nuclear transparency, defined as the
ratio between the DWIA (solid) and the PWIA (dashed) results, see eq. 
(\ref{e34}) below --- both depend on the polarization angles, but in 
different ways. First, the PWIA results depend on the polarization angles as 
a consequence of the different spatial orientation of the initial
nucleon momentum distribution, due to the different probability of finding
a given missing momentum for different orientations. In exploring
$(e,e'p)$ results of this kind, one would like to extract the
spatial orientation  of the momentum distribution, which could
be directly measured from the dashed lines if the FSI were equal to zero.

Second, the effect of the FSI (solid lines relative to dashed lines) 
is quite dependent on the polarization of the nucleus. This fact suggest that
the ``transparency'' of the nucleus to proton propagation can be
maximized or minimized by selecting  a particular polarization of the
nucleus. 
One can find a great variety of FSI effects, going from small to large
``transparency''. 
For instance, for $\theta^*=90^{\rm o}$, $\Delta\phi=45^{\rm o}$ (fig.~1),
the transparency is small $\sim 0.4$, while for for the opposite
polarization, $\theta^*=90^{\rm o}$,
$\Delta\phi=225^{\rm o}$ (fig.~2), the transparency is large $\sim 0.9$.
We also see that there are some cases, such as $\theta^*=135^{\rm o}$
and $\Delta\phi=0$, for which the transparency is bigger than one.

It is apparent from these results that the nuclear transparency 
can change drastically in going from one polarization to the
opposite, and that if one is able to understand physically the 
different behaviour seen for the FSI effects in figs.~1--2, then it could be 
possible to make specific predictions about the reaction for
future experiments. The goal of the next section is to explain 
this dependence, at least qualitatively, by making a semi-classical,
geometrical  picture based on the PWIA
of the new physics contained in this process,
and making a quantitative analysis of the results  in terms of the
propagation distance of nucleons across the nucleus.

\begin{figure}[htbp]
\begin{center}
\leavevmode
\epsfbox[120 610 502 800]{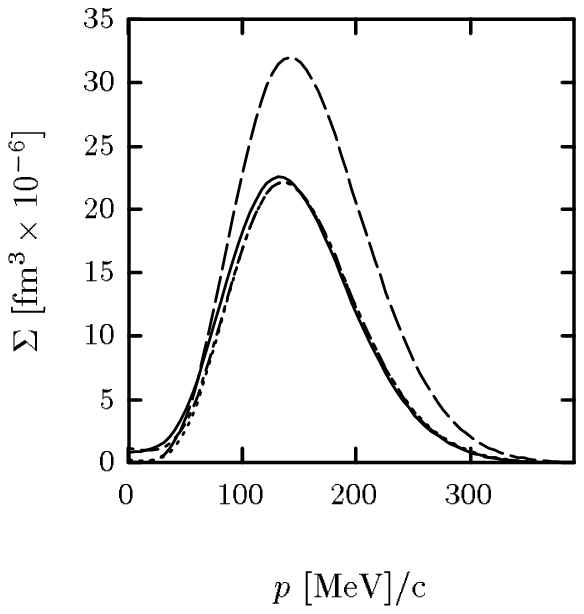}
\end{center}
{\small\sf 
{\bf Figure 3:} Unpolarized cross section. 
The meaning of the curves  is the same as in fig.~1.
}
\end{figure}
Before turning to this interpretation, let us underscore
the importance of performing experiments with polarized nuclei as compared
with the unpolarized case. This is clear if one examines fig.~3,
where we show the unpolarized cross section given as the 
average of the polarized cross section 
$\Sigma(\Omega^*)$ over all polarization angles,
\begin{equation}
\Sigma_{unpol}= \frac{1}{4\pi}\int d\Omega^* \Sigma(\Omega^*).
\end{equation}
One sees that most of the interesting behaviour 
displayed in figs.~1--2 goes away in the unpolarized case, and one is just
able to see the averaged effect of the FSI, losing the
detailed dependence of the cross section on the various polarization
directions. Importantly, with polarized nuclei more restrictions can be 
imposed both on the nuclear modeling and on the nature of the FSI effects.


\section{Polarized momentum distribution 
         and a semi-classical picture of
         the reaction}


In order to understand physically the results of the previous section in 
terms of proton propagation inside the nucleus, it is very useful to 
invoke the PWIA. Within this approximation one could try to make a 
semi-classical model of the reaction by assuming it to take place in two or more
steps as follows: first a proton with (missing) momentum $\np$ and energy
$\epsilon_{\ell j}$ is knocked-out by the virtual photon and it acquires 
momentum $\np'$ and kinetic energy $t'=\epsilon_{\ell j}+\omega$. Second, 
as this high-energy nucleon traverses the nucleus it undergoes 
elastic and inelastic scattering. In our model, the elastic scattering is 
produced by the real part of the optical potential, and the inelastic
scattering  produces  transitions of the proton into  channels other
than the elastic one, which can be  phenomenologically 
treated as absorption due to the imaginary part of the optical potential.

The important point here is that the nucleus is polarized in a 
specific direction. Accordingly, the initial-state nucleon 
can be localized in an oriented 
(quantum) orbit. From the knowledge of this orbit and of the missing momentum
one can predict the most probable location of 
the struck proton, i.e., computing  the expectation value of its position 
before the reaction takes place, and therefore one can specify the quantity
of nuclear matter that the proton must cross before
exiting from the nucleus with momentum $\np'$. 

\subsection{Distribution of a $d_{3/2}$ wave}

We illustrate the case of a particle in a $d_{3/2}$ wave, because in the
extreme shell model employed here it is the relevant state for the reaction
$^{39}\vec{\rm K}(e,e'p)^{38}{\rm Ar}_{g.s.}$, in which the residual 
$^{38}$Ar nucleus in its ground state is described as two protons 
in the $d_{3/2}$ orbit coupled to total angular momentum $J_B=0$.
For this model of the reaction it is the third proton in the $d_{3/2}$
orbit which carries the angular momentum of the initial nucleus $^{39}$K.
The wave function for a single particle in the $d_{3/2}$ shell
polarized in the $z$-direction ($\nOmega^* = \nne_3$; i.e., the particle 
is in the $|\frac32\frac32\rangle$ state), is given by
\begin{eqnarray}
\textstyle
|\frac32\frac32\rangle 
&=& 
\textstyle
\langle 21\frac12\frac12|\frac32\frac32\rangle
Y_{21}(\theta,\phi)R(r)|\uparrow\rangle +
\langle 22\frac12-\frac12|\frac32\frac32\rangle
Y_{22}(\theta,\phi)R(r)|\downarrow\rangle \nonumber\\
&=& \psi_1 |\uparrow\rangle +\psi_2 |\downarrow\rangle .
\label{e11}
\end{eqnarray}
Inserting the appropriate values for the Clebsch-Gordan coefficients
and the spherical harmonics we obtain for the spinor wave functions
\begin{eqnarray}
\psi_1 &=& -\sqrt{\frac{3}{8\pi}}\sin\theta\cos\theta\;
           {\rm e}^{i\phi}R(r) 
           \label{e12}\\
\psi_2 &=& -\sqrt{\frac{3}{8\pi}}\sin^2\theta\;
           {\rm e}^{2i\phi}R(r).
            \label{e13}
\end{eqnarray}
Here the angles $(\theta,\phi)$ are the spherical coordinates of the 
particle's position $\nr$ and $R(r)$ is its radial wave function. 
The total wave function in momentum space 
(Fourier transform) has the same angular dependence with respect to the 
angles of the missing momentum $\np$, the only differences being that the 
radial wave function is in momentum space and a global phase enters. 
The  spatial distribution is then given by the single-particle probability 
density
\begin{equation}
\rho(\nr) = |\psi_1|^2+|\psi_2|^2 =
    \frac{3}{8\pi}\sin^2\theta|R(r)|^2.
\label{e17}
\end{equation}

\begin{figure}
\begin{center}
\leavevmode
\epsfbox{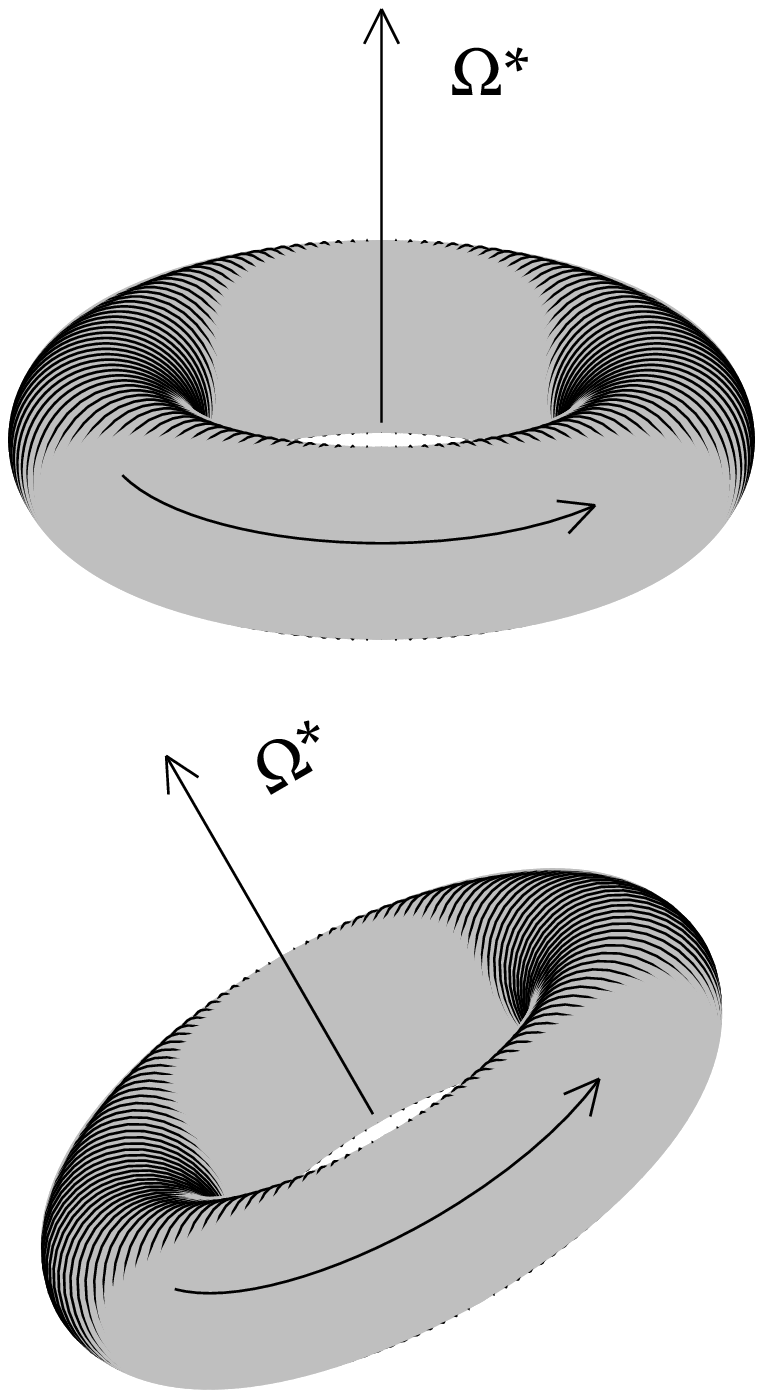} 
\end{center}
{\small\sf 
{\bf Figure 4:} 
Pictorial representation of the spatial distribution of a proton in the
$d_{3/2}$  shell, shown as a torus-like distribution for 
two different  polarizations.
}
\end{figure}

Taking into account the form of the radial wave function for the $d_{3/2}$ 
wave, we can see that the particle is distributed around the center
of the nucleus in a toroidal-like orbit as shown schematically in
fig.~4 (upper part).
In a semi-classical picture of the bound state, we can imagine the particle 
performing a rotatory orbit within the torus in a counter-clock
sense. The sense of rotation can be deduced from the direction 
of the angular momentum, namely it has to point predominantly along the
$z$-direction since the nucleon has spin-$1/2$, and the value
$J_z=\frac32$ can only be reached if the orbital angular momentum also points
in the positive $z$-direction This idea is corroborated by the model 
introduced in next section  for the local momentum of the nucleon.
The momentum distribution, which can be 
obtained using the Fourier transform of the above
wave function, has a similar shape in momentum space.
The shape of the distribution for arbitrary polarization $\nOmega^*$
is just a rotation of the above distribution, as also shown in
fig.~4 (bottom).

\subsection{Expectation value of the position for a given missing momentum}

Having established the three-dimensional shape of the nucleon orbit, 
the next step is to localize the particle within  the orbit
for a  given value  of the missing momentum $\np$.
To this end, from elementary quantum mechanics we first recall that 
the probability current for a given wave function, $\psi(\nr)$,
can be written as  
\begin{equation}
\nj(\nr) = \frac{1}{M} {\rm Re}\; \psi^{\dagger}(\nr)(-i\nabla)\psi(\nr).
\end{equation}
In a semi-classical picture from this one can define the local velocity
$\nv(\nr)$ of the particle as
\begin{equation}
\nj(\nr) = \nv(\nr)\rho(\nr)
\end{equation}
with the particle density
\begin{equation}
\rho(\nr) = \psi^{\dagger}(\nr)\psi(\nr).
\end{equation}
As a consequence, a local momentum for the particle can be defined as
the expected value of the momentum operator
$\hp = -i\nabla$ in the following way:
\begin{equation}
\np(\nr) = M\nv(\nr)= \frac{{\rm Re}\; \psi^{\dagger}(\nr)(-i\nabla)\psi(\nr)}%
                {\psi^{\dagger}(\nr)\psi(\nr)}.
\end{equation}
This equation gives us the expected value of momentum for a given
position. In order to obtain the expectation value of the {\em position}
for a given value of the {\em missing momentum} $\np$, we repeat the
above procedure, now working in momentum space. Accordingly, we employ the
Fourier transform $\tilde{\psi}(\np)$ of the wave function 
and the position operator in momentum space $\hr=i\nabla_p$ to
define the local position of the nucleon in the orbit 
for momentum $\np$ in the following way:
\begin{equation}
\nr(\np) = \frac{{\rm Re}\; \tilde\psi^{\dagger}(\np)(i\nabla_p)
                            \tilde\psi(\np)}%
                {\tilde\psi^{\dagger}(\np)\tilde\psi(\np)} .
\end{equation}
This is a well-defined vector which represents the most probable 
location of a particle with momentum $\np$ when it is described
by a wave function $\psi$. Henceforth $\nr(\np)$  represents the position
of the particle in the orbit in the present semi-classical model.

For the case of interest here of the $d_{3/2}$ orbit polarized in the
 $z$-direction, we compute the position $\nr(\np)$ by using the wave
function given in eqs. (\ref{e11}--\ref{e13}) in momentum space:
\begin{equation}
\tilde\psi^{\dagger}(\np)i\nabla_p\tilde\psi(\np)=  
\tilde\psi_1^*i\nabla_p\tilde\psi_1+
\tilde\psi_2^*i\nabla_p\tilde\psi_2 ,
\end{equation}
where now $\tilde\psi_1$, $\tilde\psi_2$ are the Fourier transforms of the 
up and down spinor wave functions. Using spherical coordinates, 
one can compute the gradient and take the real part
\begin{equation}
{\rm Re}\, \tilde\psi^{\dagger}(\np)i\nabla_p\tilde\psi(\np)=  
-\frac{3}{8\pi}\frac{|\tilde{R}(p)|^2}{p}\sin\theta(1+\sin^2\theta)\hphi ,
\end{equation}
where  $(\theta,\phi)$ are the spherical coordinates of the missing
momentum $\np$, $\tilde{R}(p)$ is the radial wave function in momentum
space, and $\hphi$ is the unit vector in the azimuthal direction.
As we see, upon dividing by the momentum distribution 
(given by eq.  (\ref{e17}), but in momentum space) 
\begin{equation}
{\tilde\psi^{\dagger}(\np)\tilde\psi(\np)}= \frac{3}{8\pi}\sin^2\theta
|\tilde{R}(p)|^2 ,
\end{equation}
the radial dependence in the numerator and denominator goes away,
and we obtain an expectation value of position which is independent
of the radial wave function --- namely, just a geometrical
quantity which is characteristic of the $d_{3/2}$ wave:
\begin{equation}
\nr(\np) = -\frac{1+\sin^2\theta}{p\sin\theta}\hphi .
\end{equation}

This expression has been obtained for the polarization direction
$\nOmega^*=\nne_3$, namely, in the $z$-direction. 
For a general polarization vector $\nOmega^*$ we just perform
a rotation of the vector $\nr(\np)$. Introducing the angle $\theta_p^*$
between $\np$ and $\nOmega^*$, 
\begin{equation}
\np\cdot\nOmega^* = p \cos\theta_p^* ,
\end{equation}
we can write the nucleon position in a way which is valid for
any polarization:
\begin{equation}
\nr(\np)=-\frac{1+\sin^2\theta_p^*}{p^2\sin^2\theta_p^*}
         \,\nOmega^*\times\np.
         \label{e23}
\end{equation}
Also, in order to illustrate the concept of momentum flow for the nucleon orbit 
introduced above, we write down the 
corresponding expression for the expectation value of momentum 
at a specific position $\nr$:
\begin{equation}
\np(\nr)=\frac{1+\sin^2\theta_r^*}{r^2\sin^2\theta_r^*}
         \nOmega^*\times\nr ,
         \label{e24}
\end{equation}
where now $\theta_r^*$ is the angle between $\nr$ and $\nOmega^*$.
Note that here there is an extra minus sign with respect to eq. (\ref{e23})
coming from the different correspondences 
$\nr\rightarrow i\nabla_p$ and $\np\rightarrow -i\nabla_r$.
Eq. (\ref{e24}) corresponds classically to a circular movement 
(orbit) around 
the rotation axis given by the polarization $\nOmega^*$.

\subsection{Applications to the $(e,e'p)$ reaction}

As a first example of the utility of the above definitions for a
physical interpretation of the results given in figs.~1--2, let us consider
the case of the $(e,e'p)$ reaction with the  $^{39}$K nucleus polarized
in the $-y$ direction ($\nOmega^*=-\nne_2$), 
given by the angles $\theta^*=90^{\rm o}$,
$\phi^*=-90^{\rm o} \Longrightarrow \Delta\phi=90^{\rm o}$.
The kinematics are illustrated in fig.~5(a).
Therein, the momentum transfer points in the $z$-direction
and we show the missing-momentum vector $\np$ corresponding to the
maximum of the momentum distribution, $p\sim 140$ MeV/c. For 
$\omega \sim 133.5$ MeV (near the quasielastic peak) the momentum 
of the ejected proton $\np'$ is also shown in the picture.
For $\nOmega^*$ pointing in the $-y$ direction, the semi-classical orbit 
lies in the $xz$-plane and follows a counter-clockwise direction of rotation.
For these conditions, the most probable 
position of the proton before the interaction is indicated with a 
black dot near the bottom of the orbit. As the particle
is going up with momentum $\np'$ after the interaction with the virtual photon, 
it has to cross all of the nucleus (not shown in the figure)
and exit it by the opposite side; thus
one expects that the FSI will be large in this situation, as shown
in the corresponding panel of fig.~1.
\begin{figure}[htb]
\begin{center}
\leavevmode
\epsfbox{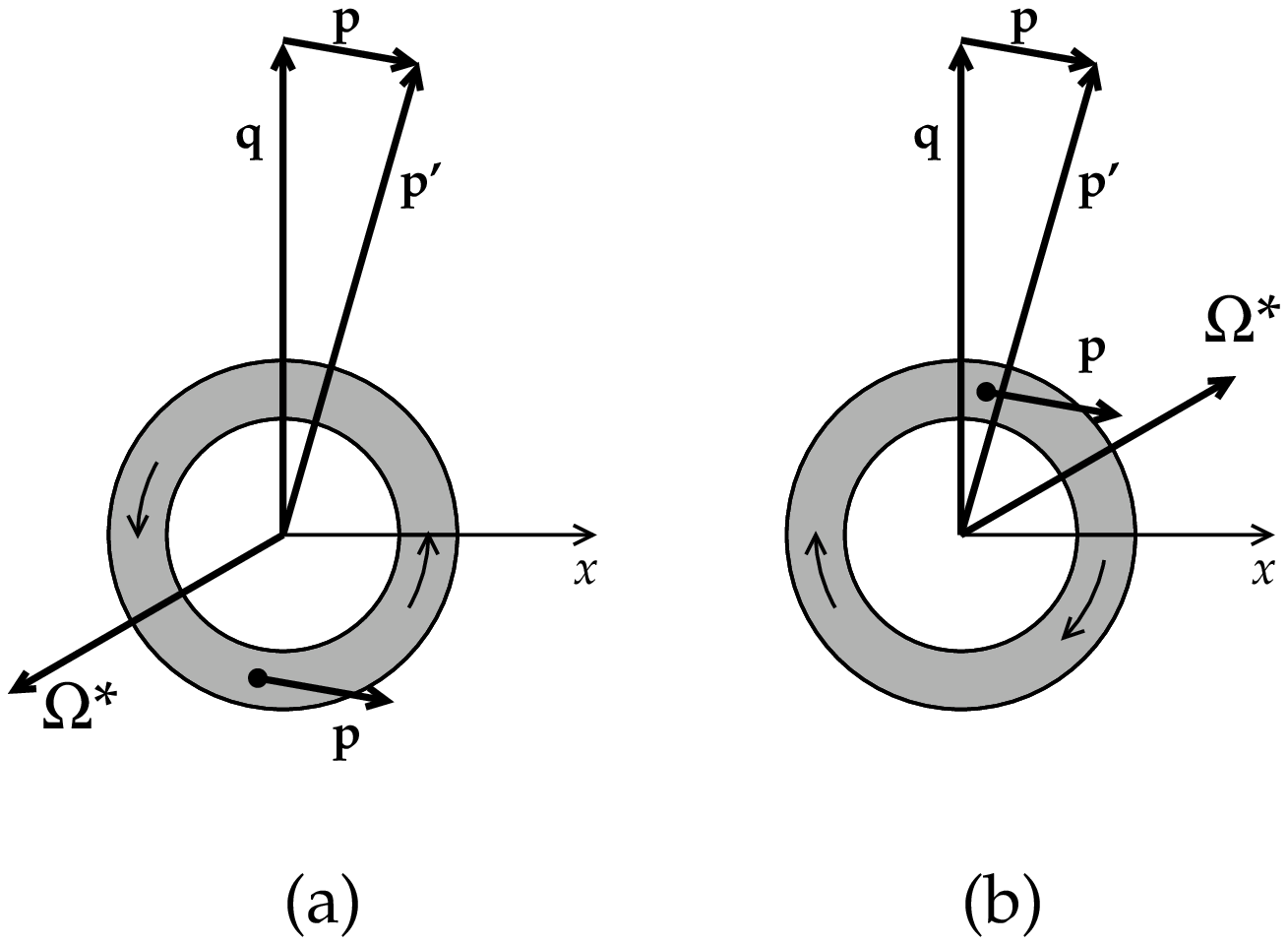}
\end{center}
{\small\sf  
{\bf Figure 5:} 
Semi-classical orbit and location of the proton for the
given kinematics and for the nuclear polarization (a) 
in the $-y$ direction, and (b) in the $y$-direction.
In the two cases the final nucleon leaves the nucleus with 
momentum $\np'$. In case (a) the nucleon is in the lower part of
the orbit and has to cross a large quantity of nuclear matter, resulting in
large FSI, whereas in case (b) the proton is in the upper part of the
orbit and crosses a small amount of matter, resulting in small
FSI.
}
\end{figure}

In fig.~5(b) we show the picture for the opposite polarization
in the $y$-direction ($\nOmega^*=\nne_2$), with angles
$\theta^*=90^{\rm o}$, 
$\phi^*=90^{\rm o}\Longrightarrow \Delta\phi=270^{\rm o}$.
In this case the nucleon distribution in the orbit is the
same as in (a), but the rotation direction is the opposite,
the sense being now clockwise. Hence now it is more probable
for the nucleon to be located near the upper part of the orbit.
As the nucleon is still going up with the same momentum $\np'$,
the distance that it has to travel through the nucleus is much smaller
than in case (a), and hence one expects small FSI effects, namely, what is
seen in the results of fig.~2.

As we can see, we have arrived at a very intuitive physical 
picture of why the FSI effects differ for different orientations of the 
nuclear spin: the polarization direction fixes the orientation of
the nucleon distribution (or in semi-classical language, the 
nucleon orbit). For a given value of the missing momentum one can 
locate the particle in a definite position within the orbit, and therefore
within the nucleus. As the particle leaves the nucleus with
known momentum $\np'$, one can immediately determine the quantity of
nuclear matter that it has to cross before exiting. 
Now we generalize the above examples to all of the polarizations
considered in this work.

\begin{figure}[hptb]
\begin{center}
\leavevmode
\epsfbox[80 400 612 780]{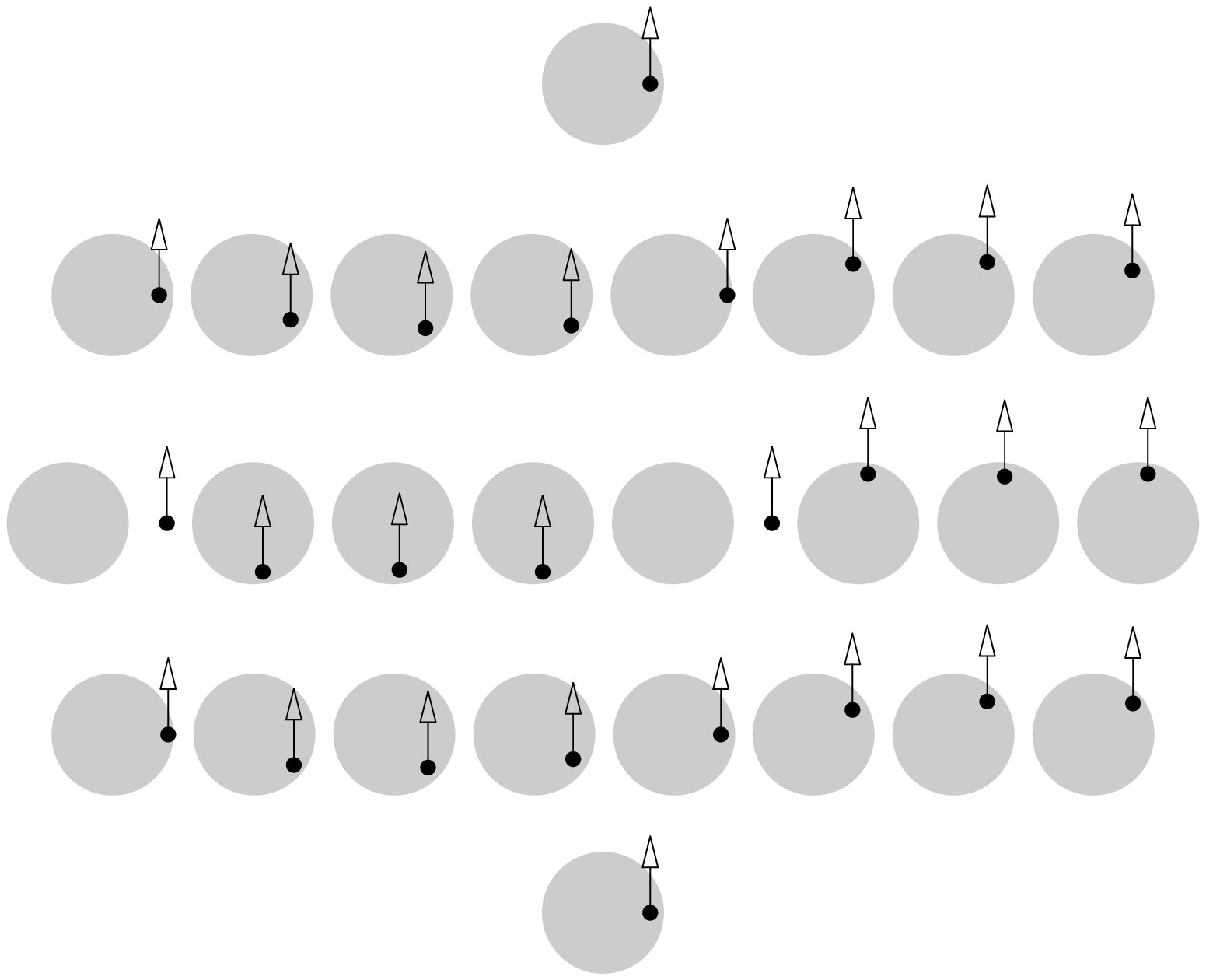}
\end{center}
{\small\sf  
{\bf Figure 6:} 
Initial position of the proton within the nucleus for each
of the polarizations of figs.~1--2 and table 1, computed 
using eq. (\ref{e23}). In each case the reference system has been
chosen so that the final momentum is going up and the position
is in the plane of the figure. In the cases where the proton is
in the lower half of the nucleus the FSI effects are found to be large 
in fig.~1, while the situations where the proton is in the upper
half, the FSI effects are small. 
The re-scattering effects produced
by the real and spin-orbit potentials are important 
only when  the proton is near the ``equator''
of the nucleus and close to its surface, 
initially with large impact parameter.
}
\end{figure}
In fig.~6 we show a general geometrical picture of the position of the 
proton before exiting the nucleus, computed using eq. (\ref{e23})
for the same kinematical conditions of figs.~1 and 2 and for all of the
26 polarizations considered in this work. The value of the missing
momentum in fig.~6 corresponds to the maximum of the cross section
(approximately located at the  maximum of the momentum distribution
$p \sim 140$ MeV/c). 

In order to make the visualization of fig.~6 clearer, we have chosen 
to have the $z$-axis in the direction of the final momentum
$\np'$; that is, in all cases the final proton is going up. 
Each circle represents the nuclear interior with radius $R\sim 3.8$ fm
and the position of the proton before the impact is shown with a black
circle. Each gray circle in fig.~6 corresponds to a different nuclear
polarization, which means a different orientation of the $d_{3/2}$
orbit and a different position $\nr$  for the nucleon. 
The $x$-axis in each one of the circles has been chosen to be in the
plane generated by the vectors $\np'$ and $\nr$ in order to present
the picture as a bi-dimensional plot. The arrows represent
the vector $\np'$. 

Each gray circle in fig.~6 corresponds to one of the 26 panels in 
figs.~1--2 and they are arranged in the same way as indicated in table 1: 
the upper and lower circles
correspond to the polarizations $\theta^*=0$ and $\theta^*=180^{\rm o}$
respectively.  Each one of the remaining three rows correspond,
from up to down, to $\theta^*=45,90,135^{\rm o}$, respectively,
while each one of the eight columns corresponds to a value of
$\Delta\phi=0,45,90,135,180,225,270,315^{\rm o}$. 
The computed values of $|\nr(\np)|$ are given in table 2.

\begin{table}
\begin{center}
\begin{tabular}{rrrrrrr} \hline
$\theta^*\,[^{\rm o}]$
      & $\Delta\phi\,[^{\rm o}]$
             & $r(\np)$ [fm] 
                       &$s$ [fm] &     $T$ & $\log T$& $\lambda$ [fm]
                                                              \\\hline
    0.&    0.&    2.799&    2.570&     .707&    -.347&    7.411\\
   45.&    0.&    2.768&    2.604&     .686&    -.376&    6.920\\
   45.&   45.&    2.712&    4.492&     .561&    -.579&    7.759\\
   45.&   90.&    2.799&    5.208&     .527&    -.640&    8.136\\
   45.&  135.&    2.970&    4.820&     .568&    -.566&    8.519\\
   45.&  180.&    3.282&    1.916&     .748&    -.290&    6.608\\
   45.&  225.&    2.970&    1.166&     .856&    -.156&    7.488\\
   45.&  270.&    2.799&    1.268&     .848&    -.165&    7.703\\
   45.&  315.&    2.712&    1.577&     .802&    -.221&    7.144\\\hline
   90.&    0.&    5.860&     .000&     .700&    -.356&     .000\\
   90.&   45.&    2.959&    6.655&     .442&    -.816&    8.155\\
   90.&   90.&    2.799&    6.552&     .445&    -.810&    8.087\\
   90.&  135.&    2.959&    6.655&     .442&    -.816&    8.155\\
   90.&  180.&    5.860&     .000&     .700&    -.356&     .000\\
   90.&  225.&    2.959&     .854&     .904&    -.101&    8.429\\
   90.&  270.&    2.799&    1.008&     .893&    -.113&    8.899\\
   90.&  315.&    2.959&     .854&     .904&    -.101&    8.429\\\hline
  135.&    0.&    3.282&    1.916&     .748&    -.290&    6.608\\
  135.&   45.&    2.970&    4.820&     .568&    -.566&    8.519\\
  135.&   90.&    2.799&    5.208&     .527&    -.640&    8.136\\
  135.&  135.&    2.712&    4.492&     .561&    -.579&    7.759\\
  135.&  180.&    2.768&    2.604&     .686&    -.376&    6.920\\
  135.&  225.&    2.712&    1.577&     .802&    -.221&    7.144\\
  135.&  270.&    2.799&    1.268&     .848&    -.165&    7.703\\
  135.&  315.&    2.970&    1.166&     .856&    -.156&    7.488\\
  180.&    0.&    2.799&    2.570&     .707&    -.347&    7.411\\ \hline
\end{tabular}
\end{center}
{\small\sf  
{\bf Table 2:} 
We show several quantities computed within our model for all
of the nuclear polarization angles considered, and for the missing
momentum at the maximum of the cross section in each 
case. From third to seventh columns we show:
the computed position of the proton $r(\np)$ within the orbit; 
the length of the nucleon path $s$ within the nucleus for nuclear
radius $R=3.8$ fm; the nuclear transparency $T$ and its logarithm
computed with just the imaginary part of the central optical
potential in the FSI; the mean free path computed as $\lambda=-s/\log T$.   
}
\end{table}

As a first qualitative analysis of the systematics of the FSI, we
can see by comparing the dashed (PWIA) and solid lines (DWIA) of
figs.~1--2, and looking at fig.~6, that in the cases where the
FSI effects are large ---predominantly in fig.~1--- the nucleon is likely 
to be located somewhere in the lower half of the nucleus, going up 
in all cases, and traversing a large amount of nuclear matter.
This happens for $\Delta\phi=45,90,135^{\rm o}$ and the most extreme
cases
correspond to $\theta^*=90^{\rm o}$, for which the proton is at the
bottom of the nucleus in fig.~6 and 
has to travel the largest distance before exiting the nucleus through
the upper surface. 

On the other hand, for $\Delta\phi=225,270,315^{\rm
o}$
(fig.~2), the nucleon is located in the upper half of the nucleus,
and it is still going up. Accordingly, the quantity of nuclear 
matter that is crossed is small and the FSI effects are also small, as seen in 
fig.~2. The most extreme cases happen again for $\theta^*=90^{\rm o}$,
where the nucleon is initially near the top of the nucleus and for
which a rather small portion of the nucleus is traversed, the FSI
effects being the smallest in fig.~2 for these cases. 

Finally, we find intermediate cases where the nucleon is somewhere 
near the ``equator'' of the nucleus, namely $\theta^*=0,180^{\rm o}$
(i.e., L and $-$L polarizations), where the FSI effects are
in between the extreme cases discussed above, and also for
$\Delta\phi=0,180^{\rm o}$, where the above semi-classical picture of 
the process is difficult to apply.

\begin{figure}[htb]
\begin{center}
\leavevmode
\epsfbox{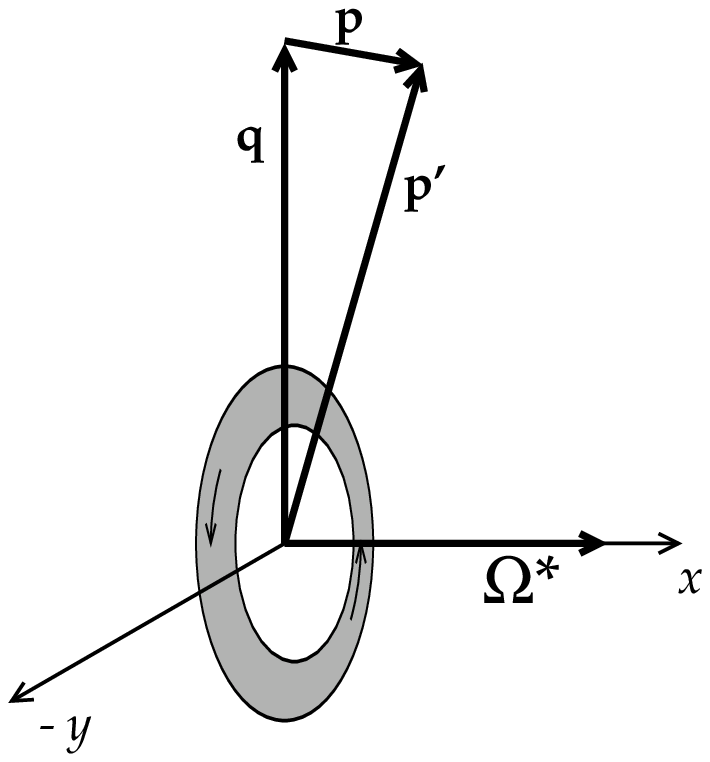}
\end{center}
{\small\sf  
{\bf Figure 7:} 
Nucleon orbit for $x$-polarization. For the present
kinematics the missing momentum is almost perpendicular to
the orbit plane, the corresponding probability of finding 
such a proton within the orbit is very small, and the FSI effects 
are relatively large.
}
\end{figure}

Consider the most extreme ``outsider''case, 
$\theta^*=90^{\rm o}$, $\Delta\phi=0$,
for which  $r(\np)=5.6$ fm and so the nucleon is ``outside'' 
of the nuclear surface (fig.~6). 
Actually what happens in this case is that the
missing-momentum vector is almost perpendicular to the plane of the 
momentum distribution, as seen in fig.~7.
In other words, the probability of finding 
a proton with momentum $\np$ is very small; so the cross section is 
also very small (less than 10\% of the cross sections for the
other polarizations shown in fig.~1) and
therefore the FSI effects are maximized for these conditions. The
reason
for the computed proton position being outside of the nucleus for this 
polarization is that the angle $\theta_p^*$ between $\np$ and
$\nOmega^*$ is very small and hence the denominator in eq. (\ref{e23})
is small, resulting in a large value of $\nr(\np)$. 

Two other interesting ``outsider'' cases correspond to the polarizations
$\Delta\phi=0$
and $\theta^*=45,135^{\rm o}$, respectively. In both cases the nucleon
near the ``equator'' of the nucleus and close to its surface (see fig.~6). 
In the first case the FSI effects are large and produce a reduction of 
the PWIA cross section by a factor $\sim 1/2$ (fig.~1). 
In the second case the FSI produce a small  {\em increase}
of the cross section (see also fig.~1). The only difference between
these two polarizations in our geometrical picture (fig.~6) is the distance
of the proton to the center of the nucleus: $r=2.77$ fm and $r=3.28$
fm respectively. As in both cases the proton is practically at
the nuclear surface, effects other than absorption are at the same
level of importance, namely, scattering by the real part
of the potential and effects relating to the spin-orbit interaction. 

In fact, the DWIA results in figs.~1--2 have been computed using an
optical potential of the type
\begin{equation}
V=V_c+V_{ls}\vec{l}\cdot\vec{s}
\end{equation}
with $V_c$ the central and $V_{ls}$ the spin-orbit part of the
potential, both being complex functions of $r$:
\begin{eqnarray}
V_c & = & U_c+iW_c \\
V_{ls} & = & U_{ls}+iW_{ls} .
\end{eqnarray}
In figs.~1--2  we also show with dotted lines the cross section
computed using only a purely absorptive central potential in the 
final state, i.e., taking $V=iW_c$ or making $U_c=V_{ls}=0$.
In contrast, the dash-dotted lines in the figures correspond to
a calculation without the spin-orbit potential, i.e., $V_{ls}=0$. 

In going from the dotted to the dash-dotted and to the solid lines
in figs.~1--2, we see that in general the major mart of the 
reduction of the cross section by the FSI is produced by the 
central absorptive part of the potential $iW_c$ (dotted lines).
The inclusion of the real part $U_c$ (dash-dotted lines) produces
a small effect in the cross section for most of the cases and the
same happens for the inclusion of the spin-orbit $V_{ls}$ (solid).

The cases having larger effects in the cross section due to the real and
spin-orbit interactions always correspond in fig.~6 to geometries
in which the initial nucleon is located near the ``equator'' and to the
right of the nucleus, i.e., initially with
large impact parameter, close to the nuclear surface and with final
momentum $\np'$ tangent to the surface. In such situations, scattering
 processes and spin-orbit interactions in the surface are maximized,
the non-absorptive part of the interaction produces deviations 
of the initial trajectory into or out of the
nucleus, and as a consequence more or less additional absorption is
produced. 

Two examples are used to illustrate this point. First, the 
$\Delta\phi=0, \theta^*=45^{\rm o}$ case shows a situation 
where the real and spin-orbit potentials produce absorption additional to
that provided by the imaginary part (fig.~1). Second, the
 $\Delta\phi=0, \theta^*=135^{\rm o}$ case is one where the nucleon is closer to
the  nuclear surface and scattering processes appear to push it
farther (fig.~1) out, so that the amount of absorption produced by the total 
central potential $V_c=U_c+iV_c$ is negligible (dash-dotted lines
in the corresponding panel). In addition, the spin-orbit interaction
produces an enhancement of the PWIA results for this polarization
(solid lines), making the total transparency bigger than one.

Clearly the surface interactions and scattering processes just discussed
vitiate the possibility of a complete  analysis of the FSI
for all of the polarizations  using a single attenuation parameter such as a
nuclear ``mean free path'' (MFP). However, the understanding gained above
does indicate how one could extract such a parameter from a selection
of the present results. 
We again start with the semi-classical concept of a nucleon orbit. 
In fact, from fig.~6 it is possible to compute the distance $s$ that the nucleon
travels across the nucleus before exiting by choosing some appropriate
value for the nuclear radius $R$. A model of exponential
attenuation of the cross section due to nuclear absorption 
(or quantum transitions to channels other than the elastic one)
can be crafted in the following way:
\begin{equation}
\Sigma_{DWIA}\simeq \Sigma_{PWIA}\;{\rm e}^{-s/\lambda} ,
\end{equation}
where $\lambda$ is the MFP and $s$ is the distance of the 
nucleon  trajectory within the nucleus.
Within this crude approximation, the nuclear transparency, defined as
the ratio between the DWIA and PWIA results, can be written as
\begin{equation}\label{e34}
T \equiv \frac{\Sigma_{DWIA}}{\Sigma_{PWIA}} \simeq {\rm e}^{-s/\lambda}.
\end{equation}

\begin{figure}[htbp]
\begin{center}
\leavevmode
\epsfbox[80 340 612 750]{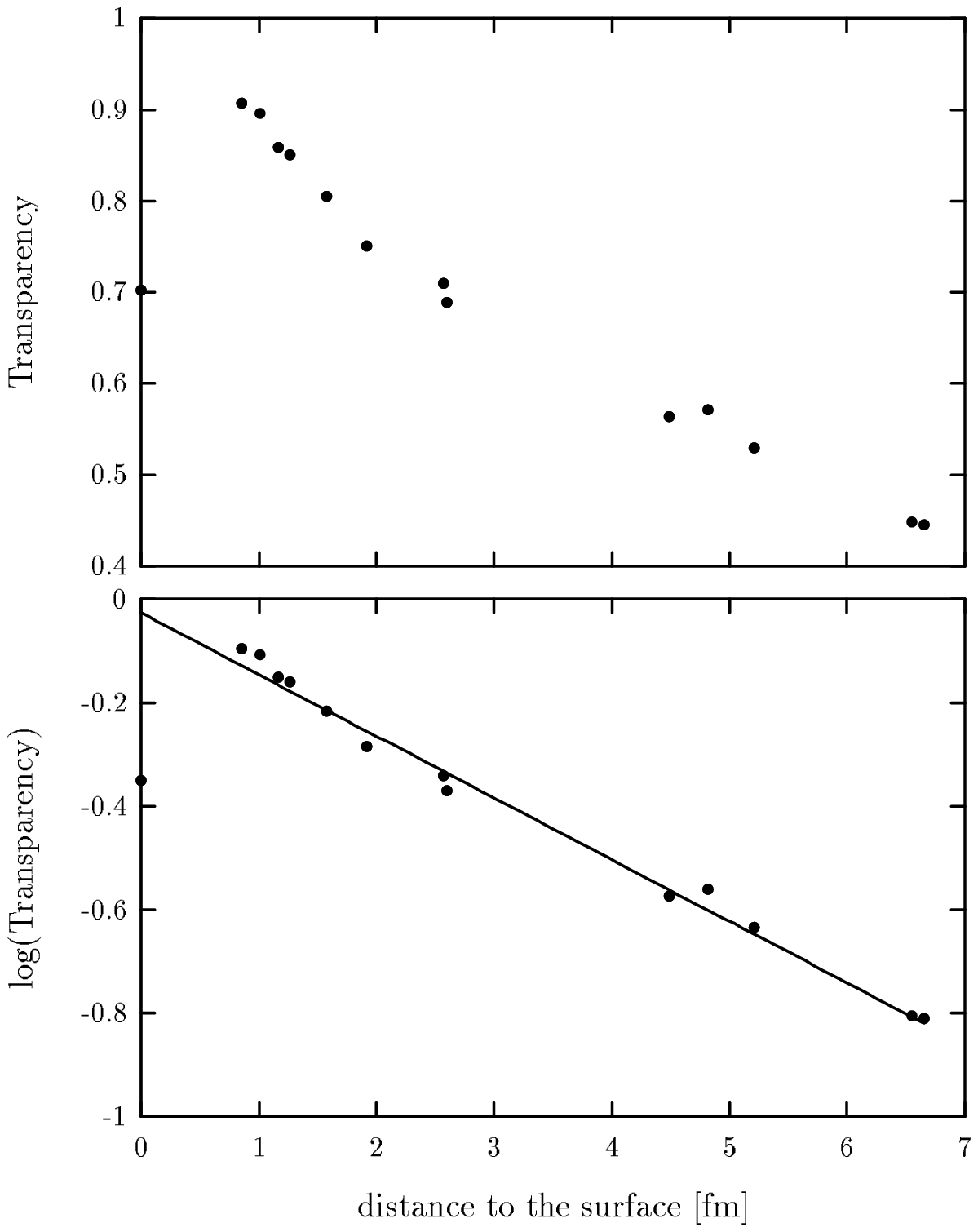}
\end{center}
{\small\sf  
{\bf Figure 8:} 
Nuclear transparency as a function of the nucleon path $s$ within the 
nucleus for the different polarizations considered in this work.
The FSI only include the imaginary part of the central optical 
potential. The $\log T$ dependence can be parametrized with a straight
line with slope $-1/\lambda$, where $\lambda$ is the mean
free path.  The point with $s=0$ is the ``outsider'' case of fig.~7, 
which cannot easily be explained within the semi-classical model.
}
\end{figure}
In fig.~8 we show the nuclear transparency as a function of the
distance $s$ to the nuclear surface, computed  for each one of the
polarizations of figs.~1--2. We have computed the transparency at the
maximum of the cross section, where for the FSI we have used just the 
imaginary part of the optical potential $V=iW_c$, without the spin-orbit 
term (i.e., we have used as $\Sigma_{DWIA}$ the dotted lines 
in figs.~1--2). The numerical values of $s$ and the
transparency are given in table 2. 

In fig.~8 we see that the dependence of $\log T$ can 
in fact be approximated by a straight line with slope
$-1/\lambda$, thus allowing one to extract an effective MFP 
in finite nuclei.

In the last column of table 2 
we also show the value of the MFP extracted from each 
polarization as $\lambda = -s/\log T$. This value is within the range
6.6 to 8.9 fm, with the exception of the two ``outsider'' cases where 
the proton is beyond the surface. This gives an averaged value of 
$\langle\lambda\rangle=7.7$ fm. 
 By performing a linear regression we obtain a MFP of
$\lambda_{lr}=8.4$ fm.  

The values for the MFP given in table 2 depend on the value chosen for
the nuclear radius $R$. The value $R=3.8$ fm chosen above corresponds
to the point $r_{1/4}$ for which the nuclear density $\rho(r)$
is 25\% of $\rho(0)$. In table 3 we present a study of the dependence
of  $\lambda$ as a function of the nuclear radius.
The length of the nucleon path through the nucleus increases with the
nuclear radius, so  
the averaged MFP increases with $R$.  For small radius
$R=r_{1/2}=3.2$ fm, the value of $\langle\lambda\rangle= 4.9$ fm,
close to the value that is used for nuclear matter. 
For this value of the radius the proton is not yet completely outside of 
the nucleus, although it is in the surface region; however, for 
finite nuclei a somewhat larger value of $R$ would be more reasonable,
since the full density of nuclear matter is not attained with finite
nuclei. In the results of table 2 
we have defined the nuclear surface as the region between 
$r_{1/2}$ and $r_{1/4}$. 

On the other hand, the values $\lambda_{lr}$ 
obtained with a linear
regression are quite independent of the radius in the
region between $r_{1/2}$ and $r_{1/10}$. Consequently we believe that 
the value of $\lambda\sim 8.5$ fm obtained in this way 
is appropriate as a ``model independent'' definition of 
the MFP for protons in finite nuclei under
the conditions of the present work. 

\begin{table}[htbp]
\begin{center}
\begin{tabular}{llll}\hline
$R$  & $\langle\lambda\rangle$
             & $\lambda_{lr}$
                    & $\rho(R)/\rho(0)$ \\\hline
 3.1  &  4.3 &  8.6 &           \\
 3.2  &  4.9 &  8.5 &   0.50 \\
 3.5  &  6.2 &  8.6 &        \\
 3.6  &  6.7 &  8.5 &   0.30 \\
 3.8  &  7.7 &  8.4 &   0.25 \\
 3.9  &  8.2 &  8.4 &   0.20 \\
 4.3  & 10.1 &  8.3 &   0.10 \\\hline
\end{tabular}\\
\end{center}
{\small\sf  
{\bf Table 3:} 
Dependence of the computed mean free path on the nuclear radius $R$.
The second column is the MFP computed for each point in fig.~8 
and then averaged. The third column is the MFP obtained with 
a linear regression of fig.~8 and the fourth column gives the
density ratio for some of the $R$-values. 
In table 2 and fig.~8 we have used the value
$R=3.8$ fm corresponding to a nuclear density which is 25\% of the
value at the origin.  
}
\end{table}

Finally, note that the above results for the MFP and nuclear 
transparency have been obtained using just the imaginary part
of the central optical potential, whereas in an experiment one cannot 
separate the different pieces of the FSI. 
The results of shown in figs.~1--2 indicate that there are situations 
where the real and spin-orbit pieces of the FSI are of
little importance in the cross section, these
situations corresponding to nuclear polarizations for which 
the impact parameter of the initial nucleon is small. Therefore,
if one considers just these cases, 
it is  still possible to extract the MFP with small error.

\section{Conclusions}

In this paper we have studied  the reaction
$^{39}$K$(e,e'p)^{38}\rm Ar_{gs}$ for polarized $^{39}$K.
The corresponding cross section has been computed within the framework of
the shell model and the FSI have been taken into account 
by using an optical potential in the final state.
The goal of the present paper has been to study the
dependence of the FSI as a function of the nuclear polarization
direction and to introduce a physical  picture of the 
results in order to understand the different effects seen 
in the cross section. 

The argument to explain the FSI effects is based on the PWIA
and it has been illustrated by introducing the semi-classical  
concept of a nucleon orbit within the nucleus. 
In fact, for given kinematics (momentum transfer, missing momentum  
and polarization angles) we  can fix the nucleon orbit {\em including its 
expected direction of motion and the expectation value of the position of 
the nucleon within the nucleus before 
the interaction.}  From this information we have computed the length
of the path that the nucleon travels across the nucleus
for each polarization.

Our results show that when the FSI effects are large the computed nucleon path 
through the nucleus is also large, whereas the opposite happens when the FSI 
effects are small. 
The importance of the real and spin-orbit pieces of the optical
potential increases with the impact parameter of the initial
nucleon with respect to the emission direction. 
Thus, by selecting the appropriate nuclear polarization, one 
can  reduce or enhance  the FSI effects and the importance 
of the real and spin-orbit pieces. Such control should prove to be 
very useful in analyzing the results from future experiments with polarized 
nuclei.

Finally, within our model we have studied the nuclear transparency as a 
function of the length of the nucleon path, providing a way of obtaining
a mean free path for protons in finite nuclei. 

In summary we have understood in some depth the role of FSI 
in $(e,e'p)$ reactions from polarized nuclei. Since by flipping the 
nuclear polarization one can go from big to small FSI effects, this
opens the possibility of using this kind of reaction to vary the roles
played by the FSI, and thus to distinguish their effects  
from other issues of interest, such as the nature of nuclear structure and 
the electromagnetic nuclear current. 

Finally, we note that in this work we have analyzed the total 
cross section for unpolarized electrons. 
It is desirable to extend this study to polarized electrons where
other spin observables and asymmetries enter. Work along these
lines is in progress. 

\section*{Acknowledgments}

The authors thank J.A. Caballero, S. Jeschonnek, R. Cenni and
A. Molinari for helpful discussions on the FSI.

This work is supported  in part by funds provided
by the U.S.  Department of Energy (D.O.E.) under cooperative agreement
\#DE-FC01-94ER40818, in part  by DGICYT   (Spain) under Contract   No.
PB92-0927 and the Junta de Andaluc\'{\i}a  (Spain) and in part by NATO
Collaborative Research Grant \#940183.


\end{document}